\documentclass{INTERSPEECH2023}
\usepackage{amsmath}
\usepackage{multirow}
\newcommand{\tabincell}[2]{\begin{tabular}{@{}#1@{}}#2\end{tabular}}
\interspeechcameraready

\title{Parameter-efficient Dysarthric Speech Recognition Using Adapter Fusion and Householder Transformation}
\name{Jinzi Qi$^1$, Hugo Van hamme$^1$}
\address{
  $^1$Department Electrical Engineering-ESAT-PSI, KULeuven, Belgium}
\email{jqi@esat.kuleuven.be, hugo.vanhamme@kuleuven.be}

\begin{document}

\maketitle
 
\begin{abstract}
In dysarthric speech recognition, data scarcity and the vast diversity between dysarthric speakers pose significant challenges. While finetuning has been a popular solution, it can lead to overfitting and low parameter efficiency. Adapter modules offer a better solution, with their small size and easy applicability. Additionally, Adapter Fusion can facilitate knowledge transfer from multiple learned adapters, but may employ more parameters. In this work, we apply Adapter Fusion for target speaker adaptation and speech recognition, achieving acceptable accuracy with significantly fewer speaker-specific trainable parameters than classical finetuning methods. We further improve the parameter efficiency of the fusion layer by reducing the size of query and key layers and using Householder transformation to reparameterize the value linear layer. Our proposed fusion layer achieves comparable recognition results to the original method with only one third of the parameters.
\end{abstract}

\section{Introduction}

The latest speech technologies, including Automatic Speech Recognition (ASR), have become increasingly popular and provide great convenience in everyday life. These technologies have traditionally focused on clear, canonical speech. However, in recent years, there has been growing interest in developing ASR models for dysarthric speech \cite{wang2021improved,geng2022speaker,tobin2022personalized,yue2022multi}, a neurological disease characterized by poor phoneme articulation. Dysarthric speakers face difficulties in daily communication, highlighting the importance of automatic dysarthric speech recognition. 

State-of-the-art ASR models \cite{zhang2020transformer, gulati2020conformer, baevski2020wav2vec} typically use a Transformer architecture \cite{vaswani2017attention} which employs millions of trainable parameters and needs hundreds of hours data for model training. However, for dysarthric speech, data scarcity has been a constant issue, due to difficulties in recruitment, collection and labeling. Thus, compared to training an E2E ASR model from scratch, finetuning an E2E ASR model \cite{shor2019personalizing, takashima2019knowledge, wang2021study} pretrained on abundant canonical speech seems more feasible. However, finetuning the entire model containing a massive number of parameters with the limited available dysarthric data can lead to over-fitting and parameter inefficiency. Moreover, dysarthric speech is influenced by mixed factors, like gender, pathogenesis (diagnosis) and severity level. The huge diversity between dysarthric speakers requires personalized model adaption \cite{geng2022speaker, tobin2022personalized, takashima2020two, turrisi2022interpretable,baskar2022speaker}. Finetuning the entire model would require a personalized model to be stored for each user, which would occupy valuable on-device storage space or require a significant upgrade in server storage for a large number of users \cite{tomanek2021residual}. 

\begin{figure}[htbp]
  \centering
  \includegraphics[width=2.5in]{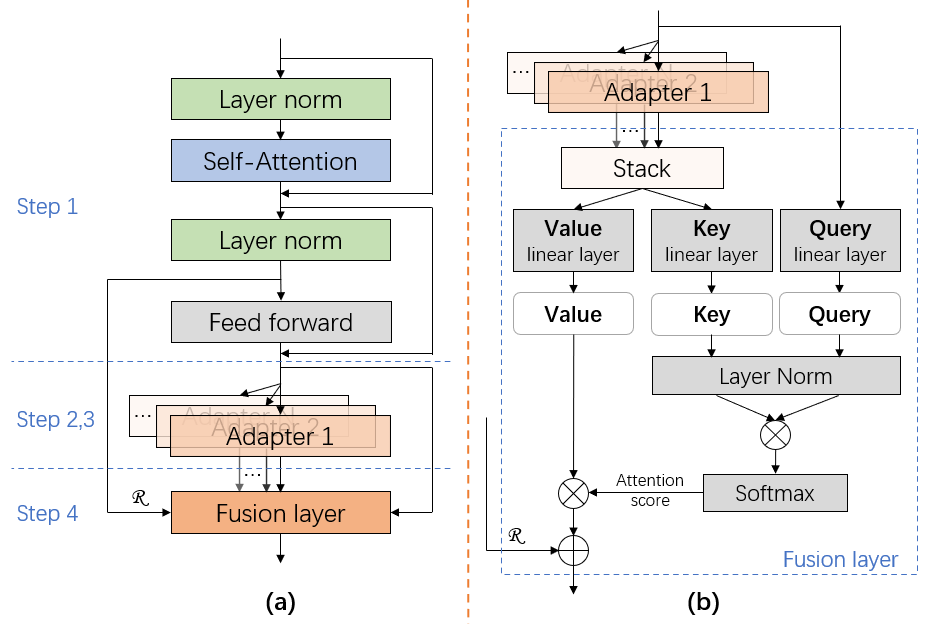}
  \caption{(a) Transformer encoder layer with adapters and fusion layer, (b) details of the fusion layer.}
  \label{fig:intro1}
\end{figure}

Adapters \cite{rebuffi2017learning, houlsby2019parameter}, which only contain a limited number of parameters, can provide a solution for both data scarcity and limited storage size. An adapter is a bottleneck module that is injected between layers of a pretrained model (figure \ref{fig:intro1}(a)). It is trained while the other parts of the pre-trained model are frozen. Previous studies \cite{kannan2019large, hou2021exploiting, karimi2021compacter, wang2021k} have demonstrated the effectiveness of Adapters in parameter-efficient transfer learning. For dysarthric speech, personalized adapters are trained and tested on atypical speech data in \cite{tomanek2021residual}, which has resulted in a similar Word Error Rate (WER) compared to finetuning. Furthermore, in \cite{baskar2022speaker}, an auxiliary net that studies speaker information is added to boost the personalized adapter performance. In this work, we use a model pretrained on canonical speech and then train the inserted adapter with dysarthric speech. 

Due to the simplicity and the small size of adapter modules, multiple adapters can be easily deployed for different groups, individuals and tasks. When trained adapters from a source domain are available, adapting to a target domain could benefit from existing adapters. To maximize the transferred knowledge among different trained adapters, Adapter Fusion \cite{pfeiffer2021adapterfusion} (figure \ref{fig:intro1}(b)) is promoted, which utilizes an attention mechanism, called fusion layer, to combine representations from source adapters and improve the performance of the target task. Specifically, the attention score is calculated and assigned to value linear layer output. For dysarthric speech, we could employ adapters trained on source speaker data and use the fusion layer to maximize the model's performance for target speaker adaptation. However, this may lead to an increase in the number of trainable parameters, which could contradict the goal of having a small storage size for personalized models. 

In this work, we apply the Adapter Fusion method to dysarthric speaker adaptation and speech recognition and investigate the feasibility of improving its parameter efficiency. Firstly, we train personalized adapters using source dysarthric speakers' data and then train the fusion layer using target speaker data for the dysarthric speech recognition task. Secondly, we trace the source of the performance improvement when using the fusion layer by ablating its two components: the attention score and the value linear layer. Furthermore, we inspect the influence of the rotation and scaling operation in the value linear layer by applying Singular Value Decomposition (SVD) to its weight matrix. Finally, we explore the possibility of improving the parameter efficiency of the fusion layer by reducing the key and query linear layer size and using Householder transformation \cite{householder1958unitary} to reformulate the rotation operation in the value linear layer. 

In section 2, we introduce the Adapter Fusion method and the application of Householder transformation. Section 3 describes the databases we used and the experimental settings. Results and analysis will be provided in section 4, and section 5 gives conclusions.

\section{Methods}

In this section, the method we use is introduced in detail. We use a encoder-decoder model with hybrid loss \cite{watanabe2018espnet} as the base ASR model.It contains a transformer encoder and two decoders: A transformer decoder and a Connectionist Temporal Classification (CTC) \cite{graves2006connectionist} decoder. For simplicity, we only insert the Adapter module in the last encoder layer and thus the fusion layer is also used in the last encoder layer. 

\subsection{Adapter Fusion method}

The transformer encoder pretrained on canonical speech $\mathcal{M}_O(\cdot)$ maps a target speaker's input sequence $\mathbf{X}$ of duration $T$, to $\mathbf{Y}_{O} = \mathcal{M}_O(\mathbf{X})$. Suppose we have $N$ dysarthric source speakers, and the personalized adapter trained by each source speaker data is $\mathcal{M}_{a_n}(\cdot)$, $n\in[1,N]$. Then the output of adapter $n$ is  $\mathbf{Y}_{a_n} = \mathcal{M}_{a_n}(\mathcal{M}_O(\mathbf{X}))$, the stacked adapter output is $\mathbf{Y}_A = [\mathbf{Y}_{a_1}, \mathbf{Y}_{a_2},...,\mathbf{Y}_{a_N}]$. 

For the fusion layer, we name the whole layer by $\mathcal{M}_{F}(\cdot)$, weight matrix of value linear layer by $\mathbf{W}$ (no bias used in this layer), the key linear layer by $\mathbf{K}(\cdot)$ and query linear layer by $\mathbf{Q}(\cdot)$. Then in the original fusion layer \cite{pfeiffer2021adapterfusion}, the fusion layer output is $\mathbf{Y}_{F} = \sum^N_{n=1} \alpha_{n}\mathbf{Y}_{a_n}\mathbf{W} + \mathcal{R}$, where $\alpha_n$ is the attention score for adapter $n$, $\alpha_n =softmax_{over~n}(\mathbf{Q}(\mathbf{Y}_O)\times\mathbf{K}(\mathbf{Y}_{a_n})')$, $\mathcal{R}$ is a residual term (see figure \ref{fig:intro1}(a)). 

To avoid gradient vanishing, different from the original version, we add a layer-normalization layer $LN(\cdot)$ for key and query term, then the attention score could be written as $\alpha_n =Softmax_{over~n}(LN(\mathbf{Q}(\mathbf{Y}_O))\times LN(\mathbf{K}(\mathbf{Y}_{a_n}))')$. In the following experiments, when we ablate the attention score in the fusion layer, the fusion output becomes $\mathbf{Y}_F = \frac{1}{N}\sum^N_{n=1} \mathbf{Y}_{a_n}\mathbf{W} +\mathcal{R}$. When eliminating the value linear layer, the fusion output is $\mathbf{Y}_F = \sum^N_{n=1} \alpha_n\mathbf{Y}_{a_n} +\mathcal{R}$.

The weight matrix of value linear layer $\mathbf{W}$ is initialized with an all-one-diagonal and the rest with small
weights ($1e-6$) \cite{pfeiffer2021adapterfusion, hou2021exploiting}. To guarantee stable adapter outputs and avoid overtraining, $\mathbf{W}$ is regularized to the identity matrix by introducing an additional loss term:
\begin{equation}
L_{reg} = ||\mathbf{I_W} - \mathbf{W}||^2 
\end{equation}
where $\mathbf{I_W}$ is an identity matrix of same size as $\mathbf{W}$. Then the total loss function during target speaker adaptation training is:

\begin{equation}
    L = (1- \lambda_1) * L_{ASR, Trans} + \lambda_1 * L_{ASR, CTC} + \lambda_2 * L_{reg}
\end{equation}
where $L_{ASR, Trans}$ is the loss from transformer decoder, $L_{ASR, CTC}$ is the loss from CTC decoder. 

If we train the fusion layer on target speaker data, the performance could benefit from fusing learned knowledge from trained source adapters. However, compared to a single personalized adapter, the fusion layer may have no advantage in the number of trainable parameters. In this work, we explore methods to improve the parameter efficiency of the fusion layer. A natural choice is to reduce the size of $\mathbf{K}(\cdot)$ and $\mathbf{Q}(\cdot)$. Further on, in the next subsection, we will work on the value linear layer.  

\subsection{Adapter Fusion with Householder transformation}

The weight matrix $\mathbf{W}$ of value linear layer acts on the adapter output. Through SVD, we obtain two orthogonal matrices $\mathbf{U}$ and $\mathbf{V}$ representing a rotation/reflection and a diagonal matrix $\mathbf{\Sigma}$ for scaling, where $\mathbf{W}=\mathbf{U\Sigma V}^T$. After training is complete, we can evaluate the effectiveness of the rotation and scaling operations on the trained $\mathbf{W}$ by forming a new matrix $\mathbf{W_{UV}}=\mathbf{UV}^T$ for rotation and another $\mathbf{W_{\Sigma}}$ for scaling. $\mathbf{W_{\Sigma}}$ is a diagonal matrix and its diagonal vector is calculated by $(\mathbf{\Sigma}_{max} - \mathbf{\Sigma}_{min})\frac{\textbf{c}-max(\textbf{c})}{max(\textbf{c})-min(\textbf{c})} + \mathbf{\Sigma}_{max}$, where $\mathbf{\Sigma}_{max}$ and $\mathbf{\Sigma}_{min}$ are the maximum and minimum of diagonal of $\mathbf{\Sigma}$, and $\mathbf{c}$ is a vector containing l-2 norm of the rows of $\mathbf{W}$.
%The latter is a diagonal matrix with $||\mathbf{W}||$ on dimension-0 as its diagonal, and is re-scaled to the value range of the diagonal of $\mathbf{{\Sigma}}$. 

Therefore, we can reformulate the matrix $\mathbf{W}$ as a scaling vector and a rotation matrix. Since the matrix is regularized to an identity matrix, there is redundancy in its parameters, and it should be full rank. To preserve the rank of the matrix while reducing the number of trainable parameters, we employ Householder transformation \cite{householder1958unitary} to reparameterize the rotation part of $\mathbf{W}$.

The Householder transformation $\mathbf{P}$ describes a reflection about a hyperplane orthogonal to $\mathbf{v}$, a length-preserving orthogonal transform :
\begin{equation}
    \mathbf{P} = \mathbf{I_W} -2\mathbf{v}\mathbf{v}^T
\end{equation}
where $\mathbf{v}$ is column unit vector. 
%Since the Householder transformation doesn't change the transferred vector's length, we can use it as one rotation operation in $\mathbf{W}$ and use multiple Householder matrices to form the entire rotation operation. 
Since the product of orthogonal matrices is again orthogonal, we can use it as a reparamerization of $\mathbf{UV}^T$.
Suppose $\mathbf{W}$ has $d$ dimensions and $d^2$ trainable elements. Then to reparameterize $\mathbf{W}$, the $\mathbf{v}$-vector should have $d$ entries, and since it's a unit vector, it has $d-1$ degrees of freedom. If we employ $d$ Householder matrices and use one scaling vector $\mathbf{s}$ ($d$-dimension), the total degrees of freedom ($d\times(d-1)+d$) would be the same as $\mathbf{W}$. If the model obtains comparable performance using less than $d$ Householder matrices, the fusion layer could achieve higher parameter efficiency. 

%To avoid the initialization on $\mathbf{v}$-vectors causing dramatic rotation on the transferred vector in the very beginning, we use $\mathbf{v}$-vectors in couple-form. A transformation matrix formed by two Householder transformations is written as:
To ensure the transform can be initialize close to the identity matrix, we use couples of $\mathbf{v}$-vectors:
\begin{equation}
\begin{split}
\mathbf{P}_c = &(\mathbf{I_W}-2\frac{(\mathbf{v}_{c,1}-\mathbf{v}_{c,2})(\mathbf{v}_{c,1}-\mathbf{v}_{c,2})^T}{||\mathbf{v}_{c,1}-\mathbf{v}_{c,2}||^2})\times\\
&(\mathbf{I_W} -2\frac{(\mathbf{v}_{c,1}+\mathbf{v}_{c,2})(\mathbf{v}_{c,1}+\mathbf{v}_{c,2})^T}{||\mathbf{v}_{c,1}+\mathbf{v}_{c,2}||^2})
\end{split}
\end{equation}
where $\mathbf{v}_{c,1}$ and $\mathbf{v}_{c,2}$ form the $c$-th couple. We initialize them with standard normally distributed values and rescale $\mathbf{v}_{c,1}$ to unit length and $\mathbf{v}_{c,2}$ to length $\frac{0.01}{\sqrt{d}}$.
%$\mathbf{v}_{c,1}$ with $\frac{randn_1}{||randn_1||}$ and $\mathbf{v}_{c,2}$ with $\frac{0.01}{\sqrt{d}}*\frac{randn_2}{||randn_2||}$, $randn_1, randn_2$ are two random sequences from standard normal distribution. 
Then the final rotation matrix $\mathbf{P}_C$ using $C$ $\mathbf{v}$-vector couples is written as: 
\begin{equation}
\begin{split}
\mathbf{P}_C = \prod_{c=1}^C\mathbf{P}_c
\end{split}
\end{equation}
Finally, we can add a diagonal scaling matrix $\mathbf{\Sigma _{s}}$ using scaling vector $\mathbf{s}$ on its diagonal:
\begin{equation}
\label{eq:WC}
    \mathbf{W}_C = \mathbf{\Sigma _{s}} * \mathbf{P}_C
\end{equation}
Notice that this does not allow to build any square matrix, but seemed to suffice for obtaining good performance.

\section{Experiments}

\subsection{Datasets}
We use \textbf{CGN} dataset \cite{oostdijk2002experiences} (excluding the ``a,c,d,e'' components) as the canonical speech data to pretrain the ASR model. It contains more than 300-hour Dutch speech. For dysarthric speech, \textbf{Domotica} dataset \cite{ons2014self} is used, which contains around 9-hour dysarthric Dutch speech from 17 speakers, in total $4173$ utterances. The content is commands related to home automation, such as ``turn on the light in the kitchen''. The intelligibility scores of speakers are provided and we categorize speakers into 3 severity levels: high (score$\textgreater$80, 5 speakers), medium (80$\ge$score$\textgreater$70, 7 speakers), low(70$\ge$score, 5 speakers). 

\subsection{Training strategy}
Training proceeds through the following steps (see figure \ref{fig:intro1}(a)):

\begin{itemize}
\item[]
\begin{description}
    \item[Step 1] Pretrain the transformer ASR model with canonical speech. Freeze the pretrained model.  
    \item[Step 2] Insert one adapter into the last encoder layer and train it with $N-3$ source dysarthric speakers' data jointly.
    \item[Step 3] For each of $N$ source dysarthric speakers, initialize the adapter from the one trained in \textbf{Step 2} and train each adapter using each source speaker's data only. Then freeze the $N$ adapters.
     %   \item[Step 3] Insert $N$ adapters into the last encoder layer of pretrained model and initialized each of them with the adapter parameters trained in \textbf{Step 2}. Train each adapter using each source speaker's data. Then freeze the adapters.    
     \item[Step 4] Insert the fusion layer after the $N$ adapters and train with the target dysarthric speaker's data.
\end{description}
\end{itemize}

In the experiment, we divide 15 dysarthric speakers into 5 subsets and each subset includes 3 speakers with different severity levels. In each trial, we use four subsets as source speakers and one subset as the target speakers. The remaining two speakers (pp34, pp35) are consistently used as source speakers. Thus number of source speakers is $N=14$. In \textbf{Step 2} we use 3 subsets of the source speakers and pp34, pp35 for training and 1 subset for validation. In \textbf{Step 3}, for each of the $N$ source speakers, we use $90\%$ data for training and $10\%$ for validation. In \textbf{Step 4}, data of each target speaker is divided into five parts, and during fusion layer training, we utilize three parts ($60\%$) for training, one for validation and one for testing in each fold. On average, $5\%$ is around $1.5$ minute per speaker and $60\%$ is about $19$ minutes. Our metric for speech recognition is Character Error Rate (CER) as it is more universal across tasks. 73 characters are used. The provided results are CER averaged over 15 speakers and five data-folds of each speaker. 

\subsection{Network and training setup}

The speech features used in the model are $83$-dimensional filter bank and pitch features. We implement the method based on the ESPnet toolkit \cite{watanabe2018espnet}. The transformer encoder has 12 layers and the transformer decoder has 6 layers. For \textbf{Step 1}, we use a batch size of $64$ and the ``Noam'' optimizer \cite{vaswani2017attention} with a learning rate of $10$ and $25000$ warm-up steps. The total number of training epochs is 230 and the final pretrained model is averaged over the 10 epochs with the highest validation accuracy. For \textbf{Step 2-4}, as well as in case of finetuning the model, we use the Adam optimizer with a learning rate of $0.001$ and early stopping with a patience of $20$. The batch size is $32$. The final model is averaged over three checkpoints with highest validation accuracy. The dimension of the transformer encoder layer output is $256$. The inner dimension (size of ``Down projection'') of the Adapter module is also chosen as $256$ since this size gives the best performance in our preliminary experiments, meaning that the module is not a traditional bottleneck shape. $\mathbf{K}(\cdot)$ and $\mathbf{Q}(\cdot)$ in the fusion layer have an original size of $256$ \cite{pfeiffer2021adapterfusion}, and we use $64$ as the reduced size. In the loss function, we set $\lambda_1 = 0.3$ and $\lambda_2 = 0.01$. Beam size $4$ is employed for joint decoding.

\section{Results}
In this section, we provide the results of using Adapter Fusion with/without Householder transformation. 

\subsection{Adapter Fusion performance}
\begin{figure}[htbp]
  \centering
  \includegraphics[width=3in]{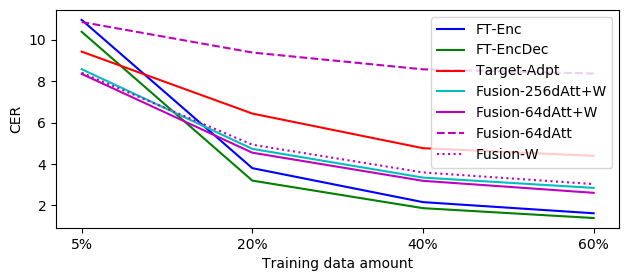}
  \caption{CER results of different models when training on different amounts of target speaker data.}
  \label{fig:1}
\end{figure}

We first compare the dysarthric speech recognition performance using different models:

\begin{itemize}
\item[]
\begin{description}
    \item[Pretrain:] Test the pretrained model obtained in \textbf{Step 1}. 
    \item[FT-Enc:] Finetune the pretrained model encoder with target speaker data. 
    \item[FT-EncDec:] Finetune the whole pretrained model with target speaker data. 
    \item[Pretrain-Adpt:] Test the model in \textbf{Step 2}. 
    \item[Source-Adpt-avg:] Test the model in \textbf{Step 3}, average outputs of all source adapters as the target speaker's output.
    \item[Target-Adpt:] Finetune the pretrained adapter with target speaker data. 
    \item[Fusion-256dAtt+W:] Test the model in \textbf{Step 4}.
    \item[Fusion-64dAtt+W:] Use fusion layer with reduced dimension 64 of $\mathbf{K}(\cdot)$ and $\mathbf{Q}(\cdot)$ in \textbf{Step 4} and test the model.
    \item[Fusion-64dAtt/W:] Use fusion layer with eliminated value linear layer / attention score in \textbf{Step 4} and test the model.    \item[Fusion-$\mathbf{W_{UV}}$/$\mathbf{W_{\Sigma}}$:] Do SVD on matrix $\mathbf{W}$ in \textbf{Fusion-W}, replace it with $\mathbf{W_{UV}}$ / $\mathbf{W_{\Sigma}}$ and test model.
\end{description}
\end{itemize}
Table \ref{tab:name1} provides the trainable parameter count and CER of these models. If the model is trained with target speaker data, the training data amounted to $60\%$ of all data. Figure \ref{fig:1} shows the CER as a function of the amount of target training data.

To assess the effectiveness of the pretrained model in speech recognition, we test it on a new canonical speech dataset \cite{K3VSND_2023} and achieved a reasonable CER of $6.3\%$. When testing on dysarthric speech, we obtained a high CER due to domain mismatch. Finetuning methods yield the best recognition results when $60\%$ data is used. However, as shown in figure \ref{fig:1}, when only $5\%$ of the data is used, the model is poorly trained with finetuning, resulting in a higher CER than other methods.

When we test the model with existing adapters, the pretrained adapter performs better than averaging source speaker adapters, as simple averaging might cause a performance drop due to the existence of individuality among adapters. Training a personalized adapter for the target speaker results in reasonable recognition results adding only $0.5\%$ parameters, while a fusion layer produces an even better CER with more parameters than \textbf{Target-Adpt}. Both methods could not surpass the finetuning methods in our experiment setting when more than $5\%$ data is used, except for the very-low-resource case ($5\%$) due to the small parameter count. By reducing the size of the $\mathbf{K}(\cdot)$ and $\mathbf{Q}(\cdot)$ to $64$, the CER is even further reduced while the layer has fewer parameters. This might be because of overtraining of the larger model.

\begin{table}[th]
\caption{Number of target-specific trainable parameters and CER in \% when using at most 60~\% of data.}
\centering
 \label{tab:name1}
\begin{tabular}{l|r|r}
\hline
\textbf{Name} & \textbf{\#para} &\textbf{CER\qquad}\\
\hline
\textbf{Pretrain} & - & 49.98 \\
\textbf{FT-Enc} & 17.7M  &1.62\\
\textbf{FT-EncDec} & 27.2M &1.39\\
\textbf{Pretrain-Adpt} & - &13.27\\
\textbf{Source-Adpt-avg} & - & 14.08\\
\textbf{Target-Adpt} & 131.5k &4.40 \\
\textbf{Fusion-256dAtt+}$\mathbf{W}$ & 197.6k& \textbf{2.85}\\
\textbf{Fusion-64dAtt+}$\mathbf{W}$ & 98.6k& \textbf{2.61}\\
\hline
\textbf{Fusion-64dAtt} & 33.0k& 8.38\\
\textbf{Fusion-}$\mathbf{W}$ & 65.5k &3.03 \\
\textbf{Fusion-}$\mathbf{W_{UV}}$ & -&6.62 \\
\textbf{Fusion-}$\mathbf{W_{\Sigma}}$ & -& 13.34\\
\hline
\textbf{Fusion-}$\mathbf{P_{64}}$ &32.8k&3.28 \\
\textbf{Fusion-}$\mathbf{W_{64}}$ &33.0k&3.19 \\
\textbf{Fusion-64dAtt+}$\mathbf{W_{64}}$ & 66.1k& \textbf{2.79} \\
\hline
\end{tabular}                       
\end{table}

To trace the performance improvement brought by the fusion layer, we ablate the value linear layer (\textbf{Fusion-64dAtt}) or the attention score (\textbf{Fusion-W}) in the fusion layer and then train it. The case \textbf{Fusion-W} gives a lower CER, indicating the greater importance of the value linear layer. This might be because the adaptation among dysarthric speakers has smaller differentiation than other tasks' adaptation \cite{pfeiffer2021adapterfusion, hou2021exploiting}. By performing SVD on the weight matrix $\mathbf{W}$ of the value linear layer, we separate it into rotation and scaling operations and evaluate the impact of each operation by testing the model \textbf{Fusion-W} with one single operation. Comparing \textbf{Fusion-}$\mathbf{W_{UV}}$ and \textbf{Fusion-}$\mathbf{W_{\Sigma}}$, we find that the rotation plays a more important role in modifying the adapter outputs $\mathbf{Y}_A$. However, even using only the scaling vector still results in a performance improvement over no action (\textbf{Source-Adpt-avg}).

\subsection{Adapter Fusion with Householder Transformation}

To enhance parameter efficiency, we aim to reformulate the rotation operation in the matrix $\mathbf{W}$ using Householder transformation. Table \ref{tab:name1} provides the number of trainable parameters and CER results using $64$ $\mathbf{v}$-vector couples to form an orthogonal $\mathbf{P}_{64}$ matrix. In \textbf{Fusion-}$\mathbf{P}_{64}$, we use $\mathbf{P}_{64}$ only as the weight matrix, achieving $3.28\%$ CER, while in \textbf{Fusion-}$\mathbf{W_{64}}$, we add the scaling as in equation (\ref{eq:WC}), which further improves the CER. In \textbf{Fusion-64dAtt+}$\mathbf{W_{64}}$, we complete the model by adding the attention, yielding a CER of $2.79\%$, which is very close to the baseline model \textbf{Fusion-64dAtt+}$\mathbf{W}$ with two-thirds of its parameters and is even higher than baseline model \textbf{Fusion-256dAtt+}$\mathbf{W}$ with only one-third of its parameters. 
%The results shows than our propose method that we reformulate the fusion layer with Householder transformation could achieve comparable dysarthric speech recognition results with higher parameter-efficiency. 

Table \ref{tab:hhtpara} compares the CER for using different $C$ values in model \textbf{Fusion-}$\mathbf{W_C}$ with the baseline case \textbf{Fusion-W}. Our results show that using $C=64$, the model \textbf{Fusion-}$\mathbf{W_{64}}$ achieves similar CER as the baseline while using only half of the parameters. Table \ref{tab:hhtpara} also demonstrates that when we have sufficient training data ($60\%$), increasing the number of $\mathbf{v}$-vectors ($C$ value) always benefits performance, and it will reach an upper limit and doesn't outperform the baseline. When training data is limited ($5\%$), increasing the $C$ value will initially improve the model performance but will then suffer from a lack of training data as well. The results show that the Householder factorization is scalable way to trade off the target-specific model size for accuracy. Notice also that applying $d$ Householder factors has a similar complexity as multiplication with a $d \times d$ matrix. 

\begin{table}[th]
\caption{CER in \% when using different $C$ values in model \textbf{Fusion-}$\mathbf{W_C}$, compared with baseline case (\textbf{bl}) \textbf{Fusion-}$\mathbf{W}$.}
\centering
 \label{tab:hhtpara}
\begin{tabular}{c|c|r|r|r|r|r}
\hline
\multirow{2}{*}{\textbf{\tabincell{c}{Training\\data \%}}} & \multicolumn{6}{c}{CER of \textbf{bl}/\textbf{Fusion-}$\mathbf{W_C}$ when \text{C} = } \\
\cline{2-7}
&\textbf{bl} & \textbf{1}&\textbf{2}&\textbf{8}&\textbf{64}&\textbf{128}\\
\hline
\textbf{5\%}  & 8.42 & 10.90 &10.48 & 9.62 & 8.90 & 9.20 \\
\hline
\textbf{60\%} & 3.03 &  7.81 &6.62 & 4.53 & 3.19 & 3.12\\
\hline
\end{tabular}                       
\end{table}

\section{Conclusions}
Dysarthric speech recognition and speaker adaptation face challenges due to data scarcity and huge diversity between dysarthric speakers. Finetuning, a common method of transferring knowledge from a rich resource domain (in our case canonical speech recognition), has drawbacks such as overfitting and high storage requirements for personalized use. Thanks to their small size and ease of use, Adapter modules offer a suitable solution. Adapter Fusion can boost knowledge transfer between learned source speaker adapters, but it may increase the number of parameters used.

In this study, we apply Adapter Fusion to target speaker adaptation for speech recognition, achieving acceptable CER results with significantly fewer trainable parameters than classical finetuning methods. We also analyze the performance improvement brought by the fusion layer and identify the critical role played by the rotation operation of the value linear layer weight matrix $\mathbf{W}$. Finally, we improve the parameter efficiency of the fusion layer by reducing the size of the query and key linear layer and reformulating $\mathbf{W}$ using Householder transformation. The proposed fusion layer achieves comparable recognition results as our starting point with only one third of the parameters.

In the future, we plan to further validate the generality of the proposed methods on additional datasets and different tasks. Additionally, we will explore solutions for the zero-shot case \cite{xu2021simple} of dysarthric speech recognition, taking practical scenarios into consideration where the model has no access to the target speakers during training.

 \section{Acknowledgements}
The research was supported by KU Leuven Special Research Fund grant C24M/22/025 and the Flemish Government under the ``Onderzoeksprogramma Artificiële Intelligentie (AI) Vlaanderen'' programme.

\bibliographystyle{IEEEtran}
\bibliography{template}

\end{document}